\documentstyle[12pt]{article}

\advance \textheight by \topskip
\begin{document}

\thispagestyle{empty}
\begin{flushright}
{\bf DFTUZ/96/08}\\
{\bf hep-th/9602106}
\end{flushright}
$\ $
\vskip 2truecm

\begin{center}

{ \Large \bf Comment on}\\
\vskip0.5cm
{\Large \bf ``New pseudoclassical model for Weyl particles"}\\
\vskip1.0cm
{ \bf Jos\'e L. Cort\'es\footnote{e-mail:
cortes@posta.unizar.es} and 
Mikhail S. Plyushchay\footnote{On leave from the
{\it Institute for High Energy Physics, Protvino,
Moscow Region, Russia}; e-mail: mikhail@posta.unizar.es}\\[0.3cm]
{\it Departamento de F\'{i}sica Te\'orica, Facultad de Ciencias}\\
{\it Universidad de Zaragoza, 50009 Zaragoza, Spain}}
\end{center}

\vskip2.0cm
\begin{center}
                            {\bf Abstract}
\end{center}
It is demonstrated that the recently proposed pseudoclassical
model for Weyl particles \cite{1} (D.M. Gitman, A.E. Gon\c calves
and I.V. Tyutin, Phys. Rev. D 50 (1994) 5439) is classically
inconsistent.  A possible way of removing the classical
inconsistency of the model is proposed.\\ [1ex] PACS number(s):
11.10.Ef, 03.65.Pm

\newpage

In a recent paper \cite{1}, Gitman, Gon\c calves and Tyutin
proposed a new pseudoclassical model to describe a Weyl
particle.  The purpose of the present comment is to demonstrate
that classically the model \cite{1}  is inconsistent.  Namely,
it will be shown that the model contains a classical relation of
the form $0=1$ being a consequence of the Lagrangian constraint,
or, equivalently, of the corresponding secondary Hamiltonian
constraint.  A possible way of removing the classical
inconsistency of the model is proposed.

Let us consider the following relation
in a superspace with real odd (Grassmann) variables
$\psi^a=\psi^{a*}$, $a=1,2,\ldots,N$,
\begin{equation}
i\sum_{a=1}^{N}\psi^a\psi^b\omega_{ab} -C\dot{=}0.
\label{1}
\end{equation}
Here $\omega_{ab}$ is some constant real antisymmetric
$c$-number matrix and
\begin{equation}
C\neq 0
\label{2}
\end{equation}
is a real constant, and the symbol $\dot{=}$ means
ordinary equality when relation (\ref{1}) is treated 
as a Lagrangian equation of motion (Lagrangian constraint)
and is a weak equality  when eq. (\ref{1}) is 
considered as a Hamiltonian constraint. 
One rewrites relation (\ref{1}) in the equivalent form
$iC^{-1}\sum_{a=1}^{N}\psi^a\psi^b\omega_{ab}\dot{=}1.$
Raising the latter condition to the 
$[\frac{N}{2}+1]$th power, $[.]$ being an integer part,
one gets on l.h.s. a sum of terms all of them containing 
factors $(\psi^{a})^2=0$. Therefore,
relation (\ref{1}) together with condition (\ref{2})
leads to the relation $0=1$, and, therefore,
it is self-contradicting.

The model of ref. \cite{1} contains the relations
of the form
\begin{equation}
T_\mu=\epsilon_{\mu\nu\lambda\rho}\pi^{\nu}\psi^{\lambda}
\psi^{\rho}+i\frac{\alpha}{2}\pi_\mu\dot{=}0,
\label{3}
\end{equation}
\begin{equation}
\pi^2=\pi_\mu\pi^\mu\dot{=}0,
\label{4}
\end{equation}
\begin{equation}
\pi_\mu\psi^\mu\dot{=}0,
\label{4*}
\end{equation}
where $\pi_\mu$ is an even canonical momentum being,
simultaneously, the energy-momentum vector of the particle,
and $\alpha=+1$ or $-1$. 
Taking into account the 
condition (\ref{4}), one supplements 
the light-like vector $\pi_\mu$ with the
tetrad components $n^{-}_\mu(\pi)$,
$n^{i}_\mu(\pi)$, $i=1,2$, $n^-\pi=1$,
$\pi n^i=0$, $n^-n^i=0$,
$n^i n^j=-\delta^{ij}$,
forming the complete oriented set:
$\pi_\mu n^-_\nu + n^-_\mu\pi_\nu
-n^i_\mu n^i_\nu=\eta_{\mu\nu}$,
$\epsilon^{\mu\nu\lambda\rho}\pi_{\mu}n^{-}_{\nu}n^{1}_\lambda
n^{2}_\rho=1$.
If, in addition, one takes into account the constraint (\ref{4*}),
the vector set of constraints (\ref{3})  
can be presented equivalently as
\begin{equation}
\pi_\mu\left(-2(\psi n^{1})(\psi n^{2})+i\frac{\alpha}{2}\right)\dot{=}0.
\label{6}
\end{equation}
From here one concludes that
the vector set of constraints (\ref{3})
is equivalent only to one constraint
\begin{equation}
-2(\psi n^{1})(\psi n^{2})+i\frac{\alpha}{2}\dot{=}0.
\label{7}
\end{equation}
Due to this fact, 
there is only one local symmetry in the model \cite{1}
in addition to the reparametrization symmetry and the local
supersymmetry. Such a symmetry is generated
by the constraint (\ref{7}).
But the relation (\ref{7})
is exactly of the form (\ref{1}),
and one arrives at the conclusion that the model \cite{1}
is classically inconsistent.

The problem connected with the constraint (\ref{3})
can be removed in the following way,
which was used earlier in refs. \cite{2}
for constructing the pseudoclassical model
for planar fermions.
Let us extend the model by introducing a pair
of scalar mutually conjugate odd variables
$\theta^+$ and $\theta^-=(\theta^+)^*$,
having the Dirac brackets
\begin{equation}
\{\theta^+,\theta^-\}_D=-i.
\label{5}
\end{equation}
The corresponding kinetic term in the Lagrangian should be
$L^\theta_{kin}=\frac{i}{2}(\theta^+\dot{\theta}{}^-
+\theta^-\dot{\theta}{}^+)$.
Then, if one replaces the $c$-number parameter $\alpha$
by $\tilde{\alpha}=\alpha\theta^+\theta^-$,
the action of model \cite{1}
will take the form
\begin{equation}
S=\int_0^1\left(-\frac{1}{2e}(\dot{x}_\mu-i\psi_\mu \chi
-\epsilon_{\mu\nu\lambda\rho}b^\nu\psi^\lambda\psi^\rho+
i\frac{\alpha}{2}\theta^+\theta^- b_\mu)^2-i\psi_\mu\dot{\psi}{}^\mu
+L^\theta_{kin}\right)d\tau,
\label{8}
\end{equation}
with Lagrange multipliers
$e$ and $b_\mu=-b_\mu^*$ (even),
and $\chi$ (odd).
The local symmetries of the system are
given by the same transformation laws of the configuration space
variables given in ref. \cite{1} and supplemented
with
the following transformation properties
for the variables $\theta^\pm$
with respect to the reparametrization,
$\delta \theta^{\pm}=\dot{\theta}{}^\pm\xi$,
supersymmetry, $\delta\theta^{\pm}=0$,
and additional symmetry transformations,
$\delta\theta^{\pm}=\mp\frac{1}{2}\alpha 
e^{-1}zb\theta^\pm\kappa$,
where $\xi(\tau)$ and $\kappa(\tau)$
are infinitesimal even parameters,
and $z_\mu=\dot{x}_\mu+\ldots$ is the `elongated' velocity appearing
in action (\ref{8}) as $z^2$.
Action (\ref{8}) leads to the same set of constraints
(\ref{3})--(\ref{4*}), but with the only substitution
$\alpha\rightarrow\tilde{\alpha}$.
Such a difference is crucial, since 
one has the relation 
$-2(\psi n^{1})(\psi n^{2}) 
+~i\frac{\alpha}{2}\theta^+\theta^-\dot{=}0$
instead of the relation (\ref{7}).
This new relation is classically
consistent and singles out a nontrivial subspace in the configuration
or phase space of the system.
Therefore, the described problems with constraint (\ref{3})
are removed now. 

In correspondence with classical relation
(\ref{5}) and $\{\psi_\mu,\theta^{\pm}\}_D=0$,
the quantum analogs of the odd variables can be realized 
as $\hat{\psi}_\mu=\frac{i}{2}\gamma_\mu\otimes\sigma_3$,
$\hat{\theta}{}^{\pm}=\frac{1}{2}\cdot 1\otimes(\sigma_1\pm i\sigma_2)$,
where one assumes that the operators $\hat{\psi}_\mu$
and $\hat{\theta}{}^{\pm}$ act on the wave functions 
$\Psi(x)$, $\Psi^t=(\psi_u,\psi_d)$, whose components $\psi_u$
and $\psi_d$ are distinguished by the $\sigma$-matrix factors
in the direct products. 
Taking for the quantum analog of the classical 
term $\theta^+\theta^-$ the same (``normal") ordering
for noncommuting operators $\hat{\theta}{}^+$ and
$\hat{\theta}{}^{-}$, we get
$\hat{\tilde{\alpha}}=\alpha\frac{1}{2}1\otimes(1+\sigma_3)$.
As a result, taking into account the quantum
analog of the constraint (\ref{4*}),
the constraint (\ref{3}) (with the described
substitution for the parameter $\alpha$),
will turn into the equation $\hat{\pi}_\mu(\gamma^5\otimes 1-
\alpha \frac{1}{2}1\otimes(1+\sigma_3))\Psi=0$
in correspondence with the classical picture analyzed above.
From here one finds that $\psi_d=0$, and
for $\alpha=+1$ and $\alpha=-1$
the component $\psi_u$ will describe the Weyl
particle of the corresponding helicity
(see ref. \cite{1}).
If, instead, one takes $\alpha=\pm 2$ and chooses 
the antisymmetrized ordering for the operators,
$\theta^+\theta^-\rightarrow \frac{1}{2}(\hat{\theta}{}^+
\hat{\theta}{}^- - \hat{\theta}{}^-\hat{\theta}{}^+)$,
both components $\psi_u$ and $\psi_d$ will survive,
and one will have a $P,T-$invariant system of Weyl
particles with opposite helicity values.

To conclude, it is worth to note that the same problems with
relations of the form (\ref{1}) are peculiar to different
related pseudoclassical models considered in literature
\cite{3}.  The corresponding models can also be improved with
the help of the extension procedure described here.
\vskip0.2cm
The work was supported by MEC-DGICYT (Spain).

\end{document}